\documentclass{article}

\usepackage{PRIMEarxiv}

\usepackage[utf8]{inputenc} 
\usepackage[T1]{fontenc}    
\usepackage{hyperref}       
\usepackage{url}            
\usepackage{booktabs}       
\usepackage{multirow}       
\usepackage{amsfonts}       
\usepackage{graphicx}
\usepackage{xcolor}
\usepackage{amsmath}
\usepackage{amssymb}
\usepackage{bm}
\usepackage[super,comma,sort&compress]{natbib}
\usepackage{lineno}
\usepackage{algorithm}
\usepackage{algpseudocode}
\floatstyle{ruled}
\restylefloat{algorithm}

\usepackage{amsthm}
\theoremstyle{definition}
\newtheorem{example}{Example}

\graphicspath{{Figures/}}

\title{PatternFormer: Learning Multiple Solution Patterns in Reaction--Diffusion Systems}

\author{
  Zhipeng Chang \\
  Department of Mathematics \\
  Penn State University \\
  University Park, PA, USA\\
  \texttt{zfc5231@psu.edu} \\
  \And
  Wenpeng Yin \\
  Department of Computer Science and Engineering \\
  Penn State University \\
  University Park, PA, USA\\
  \texttt{wenpeng@psu.edu} \\
  \And
  Wenrui Hao\thanks{*Corresponding author: Wenrui Hao (wxh64@psu.edu)} \\
  Department of Mathematics \\
  Penn State University \\
  University Park, PA, USA\\
  \texttt{wxh64@psu.edu} \\
}

\begin{document}
\maketitle

\begin{abstract}
Many nonlinear models across physics, chemistry, and biology exhibit multiple solutions for the same parameters, and capturing this entire solution set is essential for understanding pattern-forming systems. Yet existing learned surrogates are fundamentally single-valued: neural operators map each parameter to a single output, and physics-informed neural networks converge to one branch. We develop \textbf{PatternFormer} (PF), a large language model-based framework for learning the multiple solutions of nonlinear partial differential equations. By transforming unordered coexisting solutions into canonical sequences, PF produces structured solution sets in a single autoregressive pass, terminating automatically for finite families and enforcing physical residual constraints for unbounded ones. On nonlinear elliptic problems it recovers all solution branches in one inference step; on Gray--Scott it generates coexisting Turing patterns, including physically valid states absent from the reference data and beyond training. PF can also be sequentially fine-tuned across multistable systems, toward general foundation models for solution landscapes.
\end{abstract}

\vspace{0.5cm}

\section{Introduction}

Pattern formation transforms initially uniform states into complex spatial structures and is a defining feature of nonlinear systems across physics~\cite{cross1993}, chemistry~\cite{grayscott1984,pearson1993}, and biology~\cite{turing1952,gierer1972}. A fundamental property of many pattern-forming systems is multistability: nonlinear interactions, symmetry, and bifurcations allow the same parameters to support multiple coexisting steady states~\cite{wu2024,hao2020spatial}. Reaction--diffusion systems, where diffusion-driven (Turing) instability first revealed a mechanism for spontaneous pattern formation~\cite{turing1952}, provide a canonical setting for studying such phenomena, with the Gray--Scott model exhibiting particularly rich solution landscapes. In these systems, the scientific objective is often not to identify a single state, but to characterize the entire family of physically admissible states associated with a given parameter regime. Efficiently learning and exploring such solution landscapes remains a fundamental challenge.

Conventional numerical approaches face substantial computational barriers in this setting. Resolving fine-scale patterns requires high-resolution spatial discretization~\cite{dautilia2019}, while discovering multiple coexisting solutions requires repeated simulations from carefully selected initial conditions~\cite{hao2024cbmfem}, homotopy continuation methods~\cite{hao2013homotopy,hao2014bootstrapping}, or deflation techniques that systematically remove previously discovered solutions~\cite{farrell2015,charalampidis2018deflation}, including recent machine-learning variants~\cite{zhu2023neuraldeflation}. As the number of coexisting states grows and parameter spaces become higher dimensional, the computational cost increases rapidly. Machine learning offers the possibility of amortizing this cost across parameter spaces; however, existing approaches have primarily focused on learning single-valued mappings~\cite{lu2021deeponet,li2021fno} and remain unable to directly represent the full set of coexisting solutions.

Physics-informed neural networks (PINNs) minimize PDE residuals during training~\cite{raissi2019,giampaolo2022}, but on multistable systems they typically converge to one branch and fail to capture alternative solutions~\cite{zou2025}, a difficulty compounded by optimization stiffness and spectral bias~\cite{wang2021ntk,stiffpinn2020}. Extensions based on homotopy continuation~\cite{zheng2024hompinns}, generative latent-variable models~\cite{berzins2025ginn}, and Newton-based neural operators with external search strategies~\cite{hao2024nino,zhang2024psnn} still address solutions individually rather than learning the entire coexisting set. Neural operators provide efficient and discretization-independent mappings between function spaces~\cite{lu2021deeponet,li2021fno,wanghao2025leno,rao2023}, but they are single-valued by construction and therefore cannot naturally represent a parameter-to-multiple-solution relationship. Generative operators can produce multiple outputs~\cite{haitsiukevich2024,price2025gencast}, but in PDE settings these outputs typically represent uncertainty or stochastic variability rather than distinct steady states, do not determine solution multiplicity, and have not been developed for multistable pattern-forming systems. Thus, a general framework that directly learns a parameter-to-solution-set operator remains lacking.

Large language models (LLMs) provide a promising foundation for such a framework. In other scientific domains, pretrained models have demonstrated the ability to generate diverse valid objects rather than a single deterministic answer. For example, mathematical discovery systems based on language-model search generate populations of candidate programs satisfying desired properties~\cite{romeraparedes2024funsearch}. In scientific machine learning, transformer-based models have shown the ability to perform in-context operator learning from limited examples~\cite{yang2023icon,yang2023iconlm}, with theoretical results establishing their capacity to learn PDE solution operators under sufficient task diversity~\cite{cole2024icl}. These ideas have been extended to unsupervised operator learning~\cite{huang2025mfg} and generative in-context PDE prediction~\cite{serrano2025zebra,enma2025}. Moreover, pretrained transformers have demonstrated transfer beyond natural language domains, including scientific and physical applications~\cite{lu2022fpt}. The coexisting states of a parametric PDE form a low-dimensional solution manifold that deep networks approximate efficiently~\cite{chen2019manifold}, making an autoregressive transformer a natural hypothesis class for the finite family of coexisting solutions and consistent with related multi-operator PDE models~\cite{sun2025prose}. However, existing approaches remain focused on single-valued prediction or treat variability as uncertainty, rather than learning the structured set of coexisting steady states arising from physical multistability.

Here we introduce \emph{PatternFormer}(PF), a general framework for learning \emph{set-valued solution operators}. Instead of approximating a conventional mapping from parameters to a single solution, PF learns
\[
\mathcal{G}:p\mapsto\mathcal{S}(p)=\{u_1,\ldots,u_{K(p)}\},
\]
where $\mathcal{S}(p)$ represents the complete set of coexisting steady states associated with the parameter $p$. The central idea is to reinterpret solution-set generation as autoregressive sequence generation. Because coexisting states form an unordered set, PF imposes a canonical ordering based on solution norms, transforming multiple steady states into a structured sequence that can be generated by a pretrained LLM. This allows a single forward pass to produce multiple physically distinct solutions rather than collapsing to a single branch.

PF operates in a continuous latent space compatible with the pretrained LLM. Each solution field is compressed into a latent representation using an autoencoder and projected into the model embedding space through lightweight matching layers. Low-rank adaptation (LoRA)~\cite{hu2022lora} then fine-tunes the pretrained model for scientific generation without requiring conventional language prompts or discrete tokenization. Physics-based residual regularization further constrains generated states to satisfy the governing equations, enabling the discovery of physically valid solutions beyond the explicitly observed training examples.

The number of coexisting solutions can vary substantially across parameter space, requiring different generation strategies depending on the available prior knowledge. When theoretical analysis provides information about solution multiplicity, PF uses a classification mechanism to terminate generation after recovering the complete solution set. When the multiplicity is unknown, as in the Gray--Scott system, PF generates a prescribed number of candidate states, with additional solutions guided only by physics-residual constraints and subsequently validated through numerical refinement. This unified framework therefore addresses both finite-cardinality solution recovery and open-ended discovery of previously unseen patterns.

We demonstrate PF on nonlinear elliptic boundary-value problems and the Gray--Scott reaction--diffusion system, spanning one- and two-dimensional domains and both single- and multi-parameter settings. PF recovers complete families of coexisting PDE solutions in a single deterministic pass, provides high-quality initializations for classical solvers, and discovers diverse physically consistent Turing patterns beyond the training data. Beyond individual equations, the framework supports transfer across nonlinear systems through sequential adaptation, suggesting a path toward scientific foundation models capable of learning general solution landscapes rather than isolated input--output mappings.
\section{Results}

\subsection{PatternFormer: A Set-Valued Solution Learning Framework with a Pretrained LLM}

\begin{figure}[htbp]
    \centering
    \includegraphics[width=\textwidth]{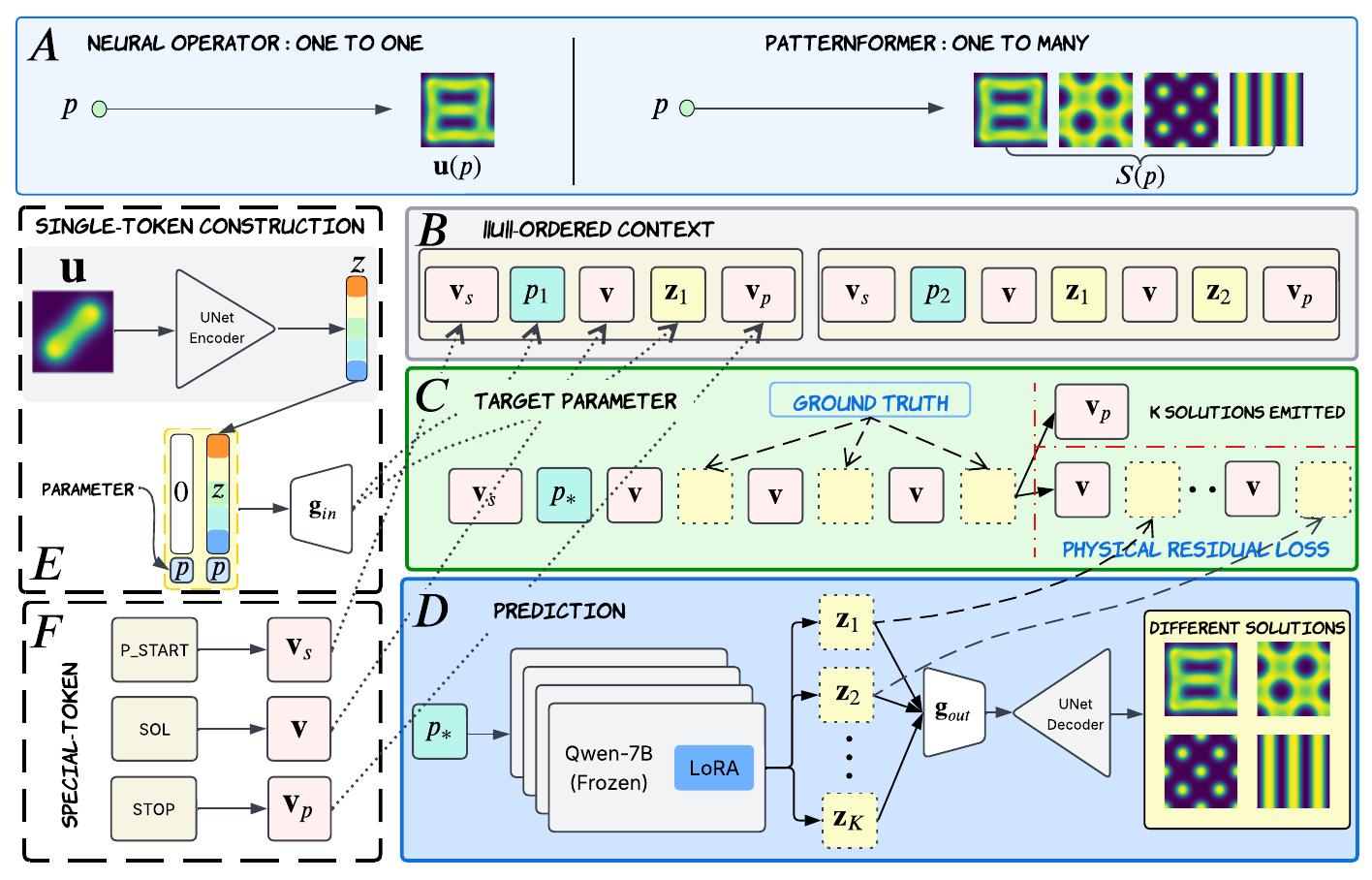}
\caption{\textbf{PatternFormer as a set-valued solution operator.}
\textbf{A}, Conventional neural operators learn a single-valued mapping from a parameter $p$ to one solution $\mathbf{u}(p)$, whereas PatternFormer (PF) learns a set-valued mapping from $p$ to the complete set $\mathcal{S}(p)$ of coexisting solutions. 
\textbf{B}, Context construction: each in-context example consists of a parameter and its complete solution set, with coexisting states ordered by solution norm and serialized into a sequence of tokens. 
\textbf{C}, Target construction: for a query parameter $p_\ast$, solutions with available reference data are supervised directly, whereas additional slots beyond the known solution count are optimized using a physics-residual loss to discover further steady states. 
\textbf{D}, Solution generation: a frozen Qwen2.5-7B model with LoRA adaptation autoregressively generates latent representations $\mathbf{z}_1,\ldots,\mathbf{z}_K$, which are decoded into distinct coexisting solutions through the output matching layer $g_{\mathrm{out}}$ and the decoder. 
\textbf{E}, Tokenization in latent space: a frozen encoder maps a solution field $\mathbf{u}$ to a latent representation $\mathbf{z}$, which is combined with the parameter $p$ and projected by the input matching layer $g_{\mathrm{in}}$ into the LLM embedding space. A parameter-only token is obtained by setting $\mathbf{z}=\mathbf{0}$. 
\textbf{F}, Special tokens: learnable embeddings represent the start, solution, and termination markers as $\mathbf{v}_{\mathrm{s}}$, $\mathbf{v}$, and $\mathbf{v}_{\mathrm{p}}$, respectively.}    \label{fig:overview}
\end{figure}

Figure~\ref{fig:overview} summarizes the architecture, which realizes the one-to-many operator $\mathcal{G}:p\mapsto\mathcal{S}(p)=\{u_1,\ldots,u_{K(p)}\}$ that single-valued neural operators cannot represent. A frozen convolutional autoencoder compresses each solution field into a compact latent representation, which is combined with the governing parameter and projected into the embedding space of the pretrained Qwen2.5-7B model~\cite{qwen2025} through a lightweight input matching layer, bypassing discrete tokenization so the LLM models continuous scientific states directly. Because coexisting states have no intrinsic ordering, we impose a canonical ordering by solution norm and generate the set autoregressively, each prediction conditioned on the parameter and the previously generated states; an output matching layer and the decoder map each latent back to a field. To handle the parameter-dependent and often unknown cardinality $K(p)$, a classification head predicts termination when an estimate of $K(p)$ is available, and otherwise the model emits a fixed number of candidates whose physics residual is minimized. These directly generated states are already near equilibrium, so a conventional solver initialized from each converges rapidly to machine precision.

\subsection{Learning complete solution landscapes of nonlinear elliptic PDEs}

We first consider the nonlinear boundary-value problem
\begin{equation}\label{eq:general-bvp}
-\Delta u + \mathcal{N}(u;p) = g(\mathbf{x}) \ \ \text{in}\ \Omega\subset\mathbb{R}^d,\qquad 
\mathcal{B}u = 0 \ \ \text{on}\ \partial\Omega,
\end{equation}
where $\mathcal{N}$ is a polynomial nonlinearity parameterized by $p$, $g$ is a source term, and $\mathcal{B}$ denotes a boundary operator. For this class of problems, analytical and numerical continuation methods provide a prior estimate of the number of coexisting solutions at each parameter. PF therefore generates the complete solution set for a given parameter and terminates autoregressive generation through its classification head once the predicted solution count is reached. Throughout, we consider only non-trivial solutions, excluding the trivial homogeneous state when present.

We study three representative problems that differ in spatial dimension, nonlinear structure, and parameterization:

\begin{example}[one-dimensional, single parameter]\label{ex:1d1p}
$d{=}1$ on $\Omega=(0,1)$, with $\mathcal{N}=u^{4}-p\,u^{2}$, $g=0$, and boundary conditions $u'(0)=0$, $u(1)=0$.
\end{example}

\begin{example}[one-dimensional, two parameters]\label{ex:1d2p}
$d{=}1$ on $\Omega=(0,1)$ with the same boundary conditions, $\mathcal{N}=a_{4}u^{4}+a_{2}u^{2}$, and $g=0$, parameterized by $p=(a_{4},a_{2})$. Example~\ref{ex:1d1p} corresponds to the slice $a_{4}=1$, $a_{2}=-p$.
\end{example}

\begin{example}[two-dimensional]\label{ex:2d}
$d{=}2$ on $\Omega=(0,1)^2$, with $\mathcal{N}=-u^{2}$, $g=-s\sin(\pi x)\sin(\pi y)$, homogeneous Dirichlet boundary conditions, and scalar parameter $p=s$.
\end{example}

The training and evaluation data are generated using classical multiple-solution numerical solvers. Examples~\ref{ex:1d1p} and~\ref{ex:2d} use the companion-based multilevel finite-element method~\cite{hao2024cbmfem}, whereas Example~\ref{ex:1d2p} uses homotopy continuation with solution maps~\cite{hao2020spatial}. Detailed problem specifications, discretizations, parameter ranges, and data splits are provided in the Supplementary Information.

Figure~\ref{fig:trackA} presents results for Example~\ref{ex:1d1p}. Across parameters with increasing solution multiplicity, from $p{=}1.62$ with one coexisting solution to $p{=}17.81$ with seven, PF generates the correct number of solution branches and accurately reconstructs their spatial profiles (Fig.~\ref{fig:trackA}A). The directly generated solutions achieve a median relative $L^2$ error of $2.3\times10^{-2}$, which is reduced to $2.9\times10^{-8}$ after a short Newton refinement requiring a median of three iterations. This refinement converges for $99.9\%$ of test parameters and maintains consistent accuracy across solution multiplicities $k=1,3,5,7$ (Fig.~\ref{fig:trackA}B).

Beyond individual solution reconstruction, PF captures the global organization of the solution landscape. The predicted bifurcation diagram, represented by $\int_0^1 u\,\mathrm{d}x$ versus $p$, reproduces the reference branch structure and the successive fold bifurcations near $p\approx4$ and $p\approx16$ (Fig.~\ref{fig:trackA}C). PF thus learns not only individual steady states but the complete family of coexisting branches and their bifurcation structure.

\begin{figure}[htbp]
    \centering
    \includegraphics[width=\textwidth,height=0.8\textheight,keepaspectratio]{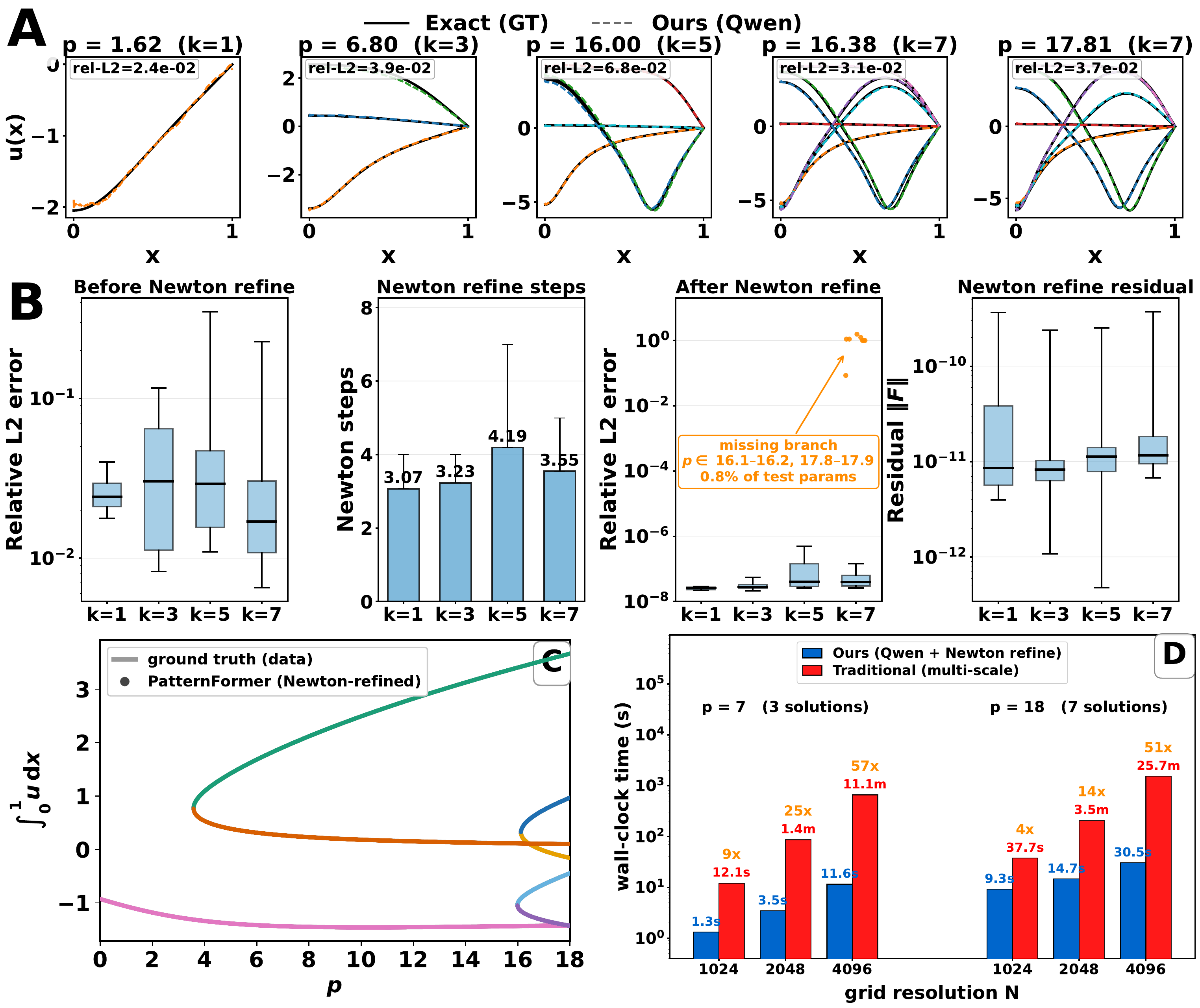}
\caption{\textbf{Learning complete solution landscapes of nonlinear elliptic PDEs.}
\textbf{A}, Representative predictions of PF for Example~\ref{ex:1d1p} at five parameter values $p$ with increasing solution multiplicity $k$. Direct model outputs (dashed) are compared with reference solutions (solid).
\textbf{B}, Relative $L^2$ errors before and after refinement, number of Newton iterations, and final residual $\lVert F\rVert$, grouped by solution multiplicity $k$. 
\textbf{C}, Recovered bifurcation structure, represented by $\int_0^1 u\,\mathrm{d}x$ versus $p$. PF reproduces the seven nontrivial solution branches (colour) and their fold bifurcations, matching the reference solution landscape (grey). 
\textbf{D}, Computational efficiency for recovering the full solution set at grid resolutions $N$ for $p=7$ and $p=18$. PF-initialized refinement (blue) is compared with a multiscale numerical solver (red). Box plots show the median and interquartile range; whiskers indicate $1.5\times$IQR.}    \label{fig:trackA}
\end{figure}

Examples~\ref{ex:1d2p} and~\ref{ex:2d} exhibit the same behavior: PF recovers the complete solution set---including, for Example~\ref{ex:1d2p}, the partition of the $(a_2,a_4)$ plane by the number of coexisting solutions---with direct predictions refining to solver precision at $99.8\%$ and $99.6\%$ convergence (per-problem errors, Newton statistics and residuals in Supplementary Fig.~S1--S2 and Supplementary Section~7).

Because PF operates in a latent representation independent of the spatial discretization, its predictions transfer across resolutions: a model trained on the $N{=}1024$ grid remains effective when interpolated to $N{=}2048$ and $4096$ without retraining (Fig.~\ref{fig:trackA}D). Used as solver initialization, these predictions accelerate recovery of the full solution set by one to two orders of magnitude at fine resolutions (per-resolution speedups in Supplementary Fig.~S1--S2 and Supplementary Section~7). PF thus turns multi-solution PDE computation from a repeated search into a single-pass generative prediction that lands within the attraction basins of distinct solution branches.
\subsection{Discovering coexisting Turing patterns in the Gray--Scott system}

\begin{figure}[htbp]
    \centering
    \includegraphics[width=\textwidth,height=0.8\textheight,keepaspectratio]{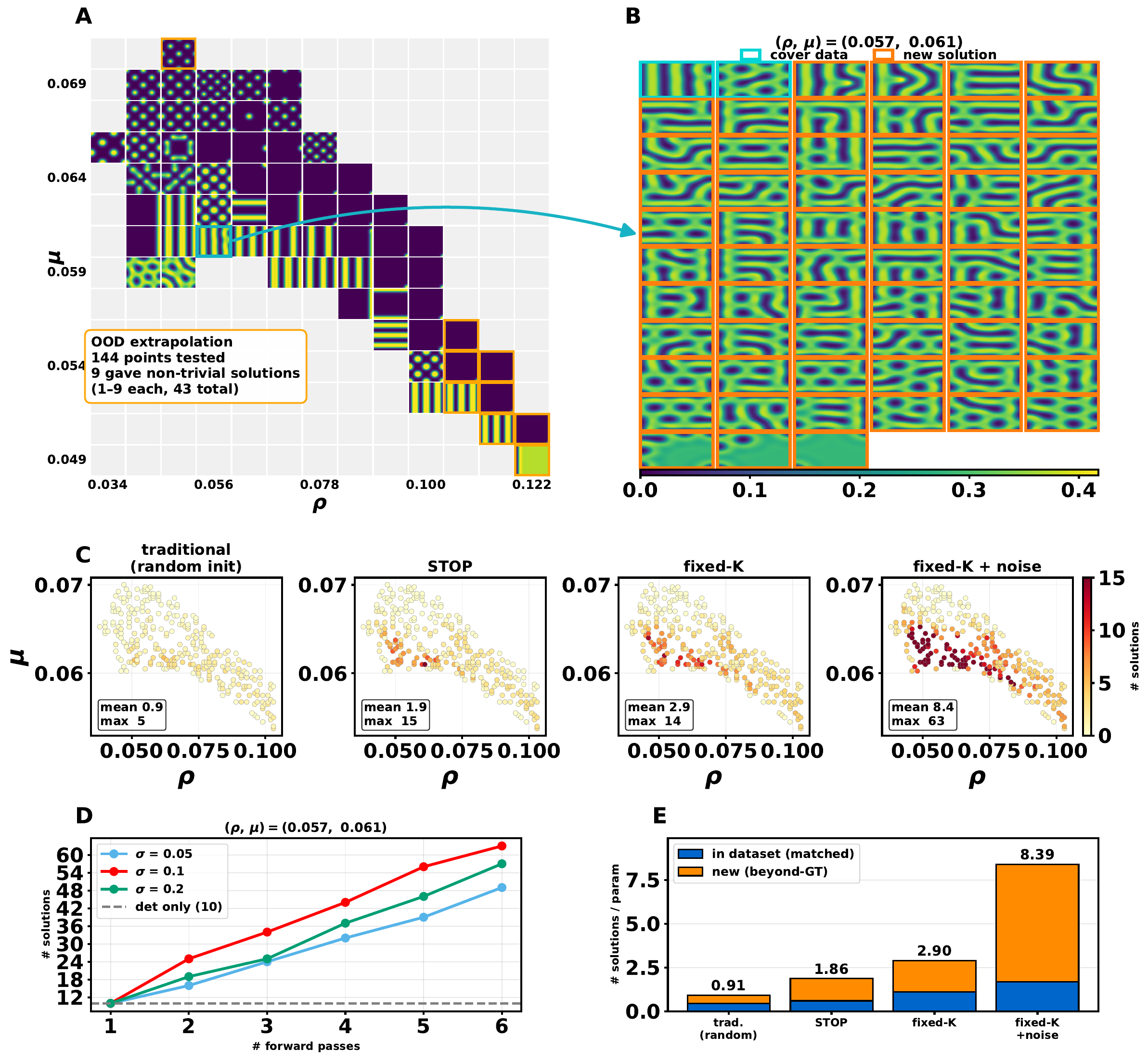}
\caption{\textbf{Open-ended discovery of coexisting Turing patterns in the Gray--Scott system.}
Statistics are computed over held-out test parameters at $D_A{=}6.25\times10^{-5}$ and $D_S{=}1.25\times10^{-4}$.
\textbf{A}, Pattern multiplicity landscape over the $(\rho,\mu)$ parameter plane. Each cell shows the largest number of distinct steady patterns discovered by PF; orange cells indicate out-of-distribution parameter regions, whereas grey cells contain only the homogeneous state.
\textbf{B}, Distinct coexisting patterns generated at the representative parameter marked by the cyan cell in \textbf{A}. Combining deterministic and noise-augmented generations, PF discovers $63$ distinct steady states, including $2$ patterns matching the reference data (cyan) and $61$ previously unseen patterns (orange). Each pattern satisfies the discretized Gray--Scott equations with an FDM residual below $10^{-9}$.
\textbf{C}, Comparison of the number of recovered solutions per parameter for four approaches: a traditional random-start solver, a stop-token design, the fixed-$K$ PF generator, and fixed-$K$ generation with noise augmentation.
\textbf{D}, Effect of latent perturbations on open-ended discovery: cumulative number of distinct solutions found as a function of forward passes for different noise levels $\sigma$ at the representative parameter.
\textbf{E}, Mean number of discovered solutions per parameter, separated into patterns matching the reference dataset and previously unseen solutions.}    \label{fig:gs}
\end{figure}

Gray--Scott provides a more challenging test in which the number of coexisting steady states is not known a priori. The steady states satisfy
\begin{equation}
D_A\,\nabla^2 A + S A^2 - (\mu+\rho)\,A = 0,\qquad
D_S\,\nabla^2 S - S A^2 + \rho(1-S) = 0
\end{equation}
on $[0,1]^2$ with Neumann boundary conditions, where $A$ and $S$ denote the activator and substrate fields, $(\rho,\mu)$ are the feed and kill rates, and $D_A,D_S$ are the diffusion coefficients. Unlike the nonlinear elliptic problems above, no complete solution enumeration is available for this system. Training and evaluation patterns are therefore generated using a quasi-Newton tensor-product solver with parameter continuation~\cite{hao2025quasinewton} (Methods; Supplementary Section~4). Here the goal is not to recover a predetermined solution count, but to generate diverse, physically valid steady states at each parameter.

At diffusion coefficients $D_A{=}6.25\times10^{-5}$ and $D_S{=}1.25\times10^{-4}$, Figure~\ref{fig:gs} evaluates PF on held-out parameters and on regions beyond the training domain. Across the $(\rho,\mu)$ plane, PF generates distinct coexisting steady states, with representative patterns summarized in Fig.~\ref{fig:gs}A. Remarkably, parameters outside the training region (orange cells) also produce valid patterns, demonstrating extrapolation beyond the observed parameter range.

To explore open-ended pattern discovery, we aggregate multiple noise-perturbed forward passes after refinement and deduplication (fields distinct when their pairwise relative $L^2$ exceeds $0.15$; all satisfy the discretized Gray--Scott equations to a finite-difference residual below $10^{-9}$, rather than being numerical artifacts). At a representative high-multiplicity parameter, PF discovers $63$ distinct coexisting patterns (Fig.~\ref{fig:gs}B), two present in the reference dataset and $61$ previously unseen. The model is thus not limited to memorizing the observed set; the physics-residual objective enables exploration of additional physically consistent patterns, even at parameters where reference solutions are already available (Supplementary Fig.~S8).

Figure~\ref{fig:gs}C,E compare the number of discovered solutions per parameter across four approaches. The deterministic fixed-$K$ generator recovers $2.90$ distinct solutions per parameter in a single forward pass, exceeding both a random-start quasi-Newton solver~\cite{hao2025quasinewton} ($0.91$) and the stop-token design ($1.86$). Adding latent perturbations turns generation into a controllable exploration mechanism: the number of recovered states grows with the number of forward passes, reaching approximately $63$ at a representative parameter after six passes (Fig.~\ref{fig:gs}D) and raising the test-set average to $8.39$ per parameter (Fig.~\ref{fig:gs}E), with the additional discoveries consisting primarily of previously unseen patterns. A second diffusion regime ($D_A{=}2.5\times10^{-4}$, $D_S{=}5\times10^{-4}$) exhibits the same behavior (Supplementary Fig.~S7).
\subsection{Extrapolation beyond the training region}

We next examine whether PF can generalize beyond the parameter regions encountered during training. Each trained model is fixed and queried at extrapolated parameters, where it directly generates candidate coexisting solutions in a single forward pass. The generated fields are subsequently refined by Newton's method at the queried parameter. Reference solution sets at these extrapolated parameters are independently computed using classical numerical solvers based on parameter continuation or multi-start searches (Supplementary Section~10). We evaluate extrapolation performance by measuring solution coverage, defined as the fraction of reference branches recovered by the model, and by assessing whether the generated candidates contain spurious solutions.

\begin{figure}[htbp]
    \centering
    \includegraphics[width=\textwidth,height=0.8\textheight,keepaspectratio]{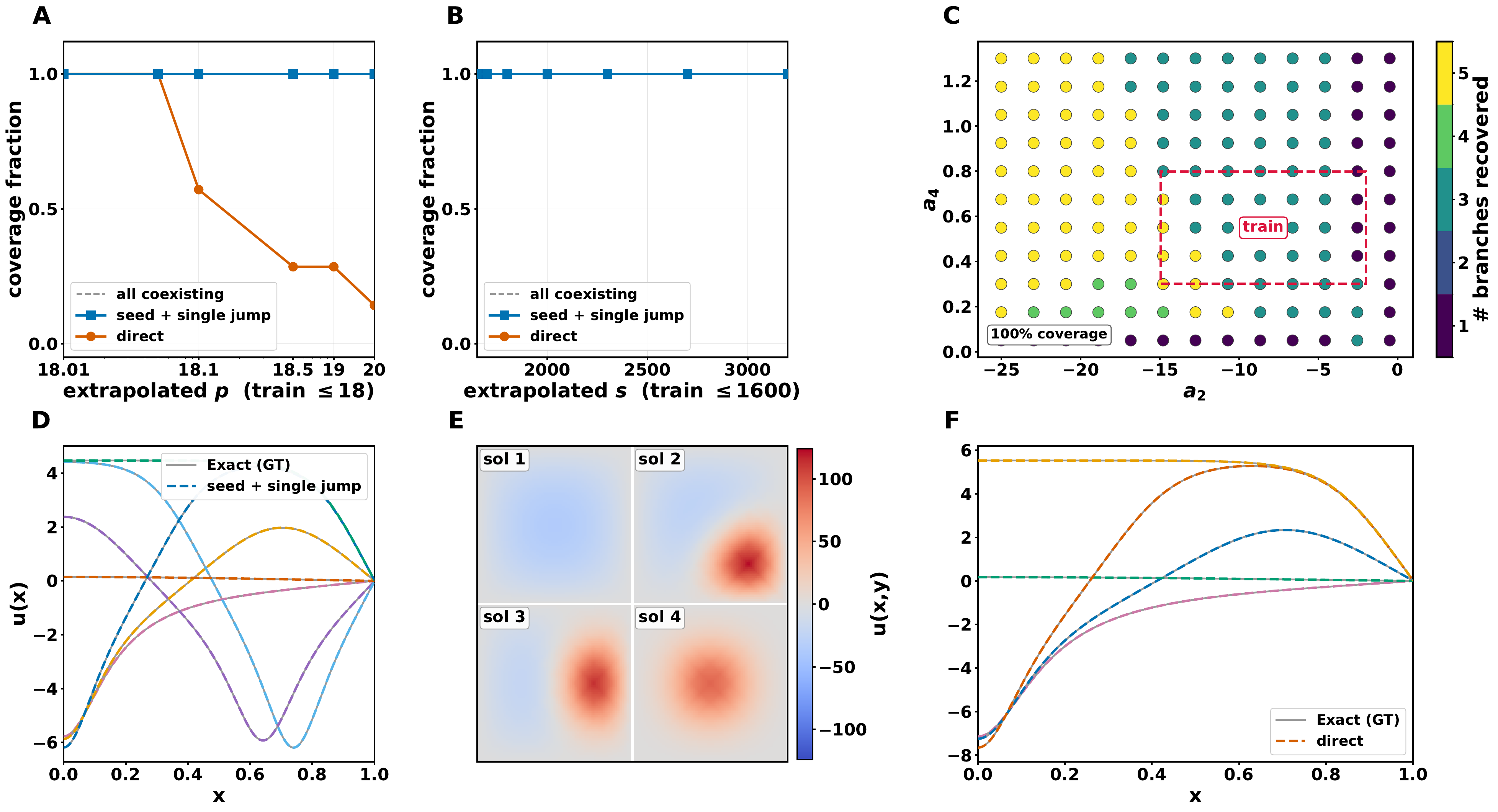}
\caption{\textbf{Extrapolation beyond the training parameter range.}
\textbf{A,B}, Recovery of reference solution branches as parameters move beyond the training domain for Example~\ref{ex:1d1p} (\textbf{A}, training region $p\leq18$) and Example~\ref{ex:2d} (\textbf{B}, training region $s\leq1600$). Direct PF generation at the extrapolated parameter (orange) is compared with Newton refinement initialized from the boundary solution set (blue); dashed lines indicate complete branch recovery. For Example~\ref{ex:2d}, both approaches maintain full coverage.
\textbf{C}, Extrapolation across the $(a_2,a_4)$ parameter plane for Example~\ref{ex:1d2p}. Each dot is a tested parameter, coloured by the number of coexisting branches recovered by the PF-seeded continuation; recovery is complete at every parameter (100\% coverage), far beyond the training box (dashed). Direct-generation and single-step results are in Supplementary Fig.~S5.
\textbf{D}, Recovery of all seven solution branches for Example~\ref{ex:1d1p} at $p=20$, comparing PF-initialized refinement (dashed) with reference solutions (solid).
\textbf{E}, Representative coexisting steady states recovered for Example~\ref{ex:2d} at an extrapolated parameter value.
\textbf{F}, Direct PF predictions for Example~\ref{ex:1d2p} at an extrapolated parameter point (dashed) compared with reference solutions (solid).}    \label{fig:extension}
\end{figure}

For the three problems with theoretically characterized solution counts (Examples~\ref{ex:1d1p}--\ref{ex:2d}), PF produces no observed false-positive branches: generated states consistently converge under Newton refinement to valid solutions of the governing equations. Direct generation alone recovers the complete solution set near the boundary of the training region, while the achievable extrapolation range depends on the problem and the direction of parameter extension (Fig.~\ref{fig:extension}A,B and Supplementary Fig.~S5).

Direct generation is not the only way to exploit PF beyond its training domain. Using its predicted solution set at the nearest in-distribution parameter as a learned initialization for Newton refinement, a \emph{single-jump} strategy recovers all seven branches of Example~\ref{ex:1d1p} at $p=20$ in about four iterations per branch (Fig.~\ref{fig:extension}A,D). Chaining this initialization through intermediate parameters (a \emph{stepped march}) transports the full branch structure much further---to $p=100$ for Example~\ref{ex:1d1p} (more than five times the largest training parameter) and across every direction of the $(a_2,a_4)$ plane for Example~\ref{ex:1d2p} (Fig.~\ref{fig:extension}C; Supplementary Fig.~S5--S6, with algorithmic details in Supplementary Section~10).
Representative extrapolated solutions for the two-dimensional and 1D two-parameter problems are shown in Fig.~\ref{fig:extension}E,F.

PF also exhibits direct extrapolation in the Gray--Scott system, where the objective is open-ended pattern discovery rather than branch coverage. Outside the training rectangle in the $(\rho,\mu)$ plane, PF generates non-trivial coexisting patterns in $9$ of $144$ tested out-of-distribution parameter cells, with each such cell containing $1$--$9$ distinct patterns (Fig.~\ref{fig:gs}A, orange cells).
The remaining cells correspond to parameter regimes where no steady patterns were identified by the reference solver. Thus, extrapolation does not lead to spurious solutions: generated states outside the training domain remain consistent with the governing equations and the observed pattern-forming regimes.

\subsection{Separating solution discovery from numerical refinement}

\begin{figure}[htbp]
    \centering
    \includegraphics[width=\textwidth,height=0.8\textheight,keepaspectratio]{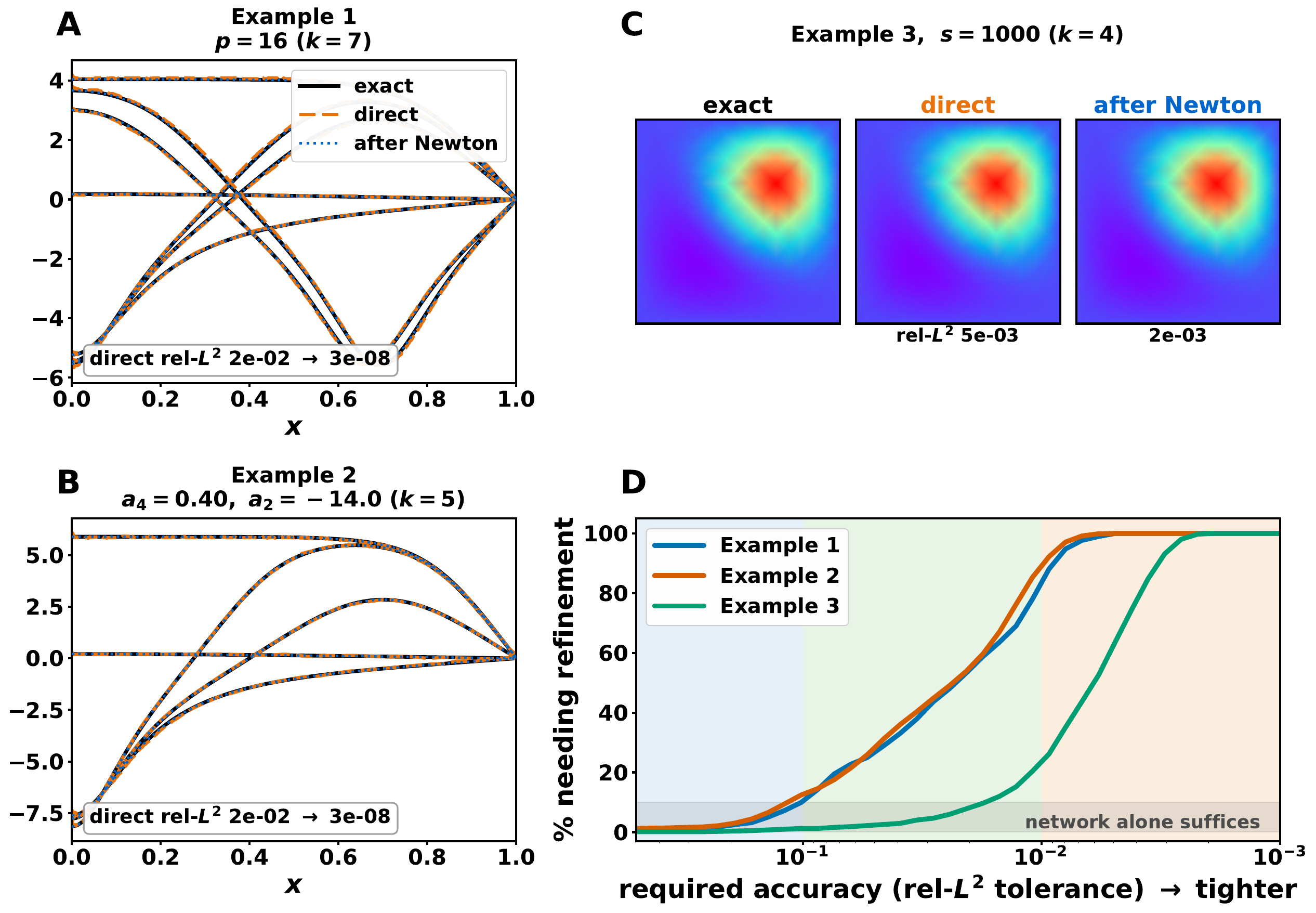}
    \caption{\textbf{PatternFormer discovers solution branches; refinement improves numerical precision.}
    \textbf{A,B}, For Example~\ref{ex:1d1p} (\textbf{A}) and Example~\ref{ex:1d2p} (\textbf{B}), direct PF predictions (orange dashed) and Newton-refined solutions (blue dotted) compared with reference solution branches (black solid). The direct predictions already recover the complete branch structure, with annotated median relative $L^2$ errors before and after refinement. 
    \textbf{C}, Example~\ref{ex:2d} at $s=1000$: reference, direct, and refined solutions for a representative branch, with per-field relative $L^2$ errors shown. 
    \textbf{D}, Accuracy-dependent refinement requirement: fraction of predictions requiring refinement as a function of the target relative $L^2$ tolerance. Refinement becomes necessary primarily when higher numerical precision is demanded.}
    \label{fig:tolerance}
\end{figure}

A central question is whether PF discovers the solution landscape or whether the subsequent numerical refinement performs the essential search. Three observations from the same experiments distinguish these roles (Fig.~\ref{fig:tolerance}). First, the direct model outputs already recover the complete set of coexisting branches, with only small deviations from the reference solutions that preserve the correct branch identity and spatial structure (Fig.~\ref{fig:tolerance}A--C); the median direct relative $L^2$ errors ($6\times10^{-3}$ to $2\times10^{-2}$) reflect local approximation error rather than missing or incorrect solutions.

Second, refinement improves precision but does not alter the discovered set: each prediction converges to the nearby PF-identified branch in three to four Newton iterations, reducing the error by several orders of magnitude without branch switching. Third, varying the requested accuracy shows that only a small fraction of predictions need refinement at moderate tolerances, refinement becoming important only near high numerical precision (Fig.~\ref{fig:tolerance}D). PF thus performs the essential discovery of coexisting steady states, while numerical refinement is an optional mechanism for solver-level accuracy.

\subsection{Pretraining and warm-starting enhance solution multiplicity recovery}

\begin{figure}[htbp]
    \centering
    \includegraphics[width=\textwidth,height=0.82\textheight,keepaspectratio]{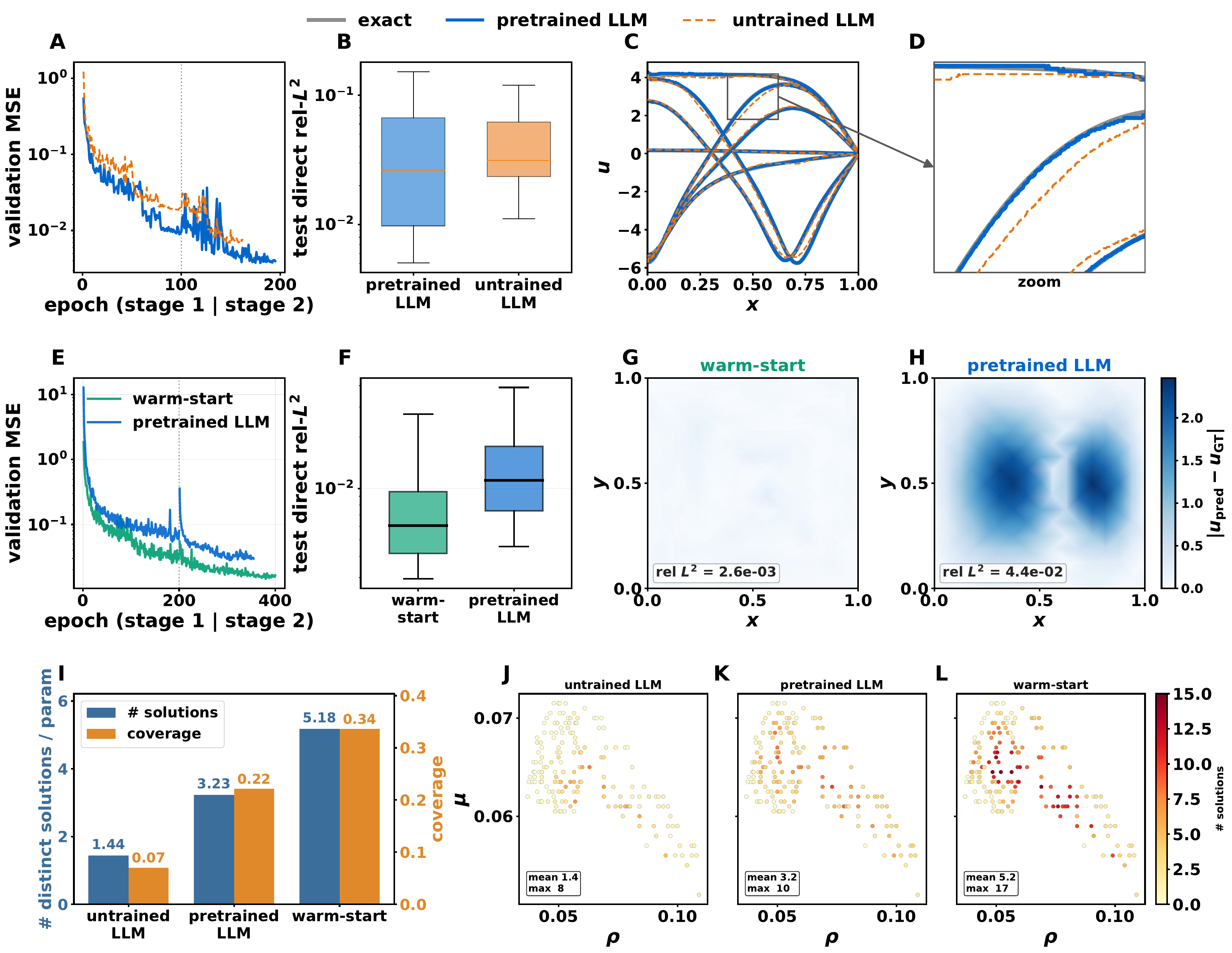}
    \caption{\textbf{Pretraining and cross-equation warm-starting improve recovery of coexisting solutions.}
    \textbf{A--D}, Example~\ref{ex:1d1p}: deterministic PatternFormer fine-tuned from pretrained Qwen2.5-7B compared with the same architecture initialized from random weights. 
    \textbf{A}, Validation MSE during the two training stages. 
    \textbf{B}, Test direct relative $L^2$ error. 
    \textbf{C,D}, Recovery of seven coexisting branches at a representative parameter, with exact solutions shown in grey and predictions from pretrained and random-weight models compared. 
    \textbf{E--H}, Cross-equation warm-starting from Example~\ref{ex:1d2p} to Example~\ref{ex:2d}. Warm-started training (green) is compared with a fresh adapter trained from the pretrained Qwen initialization (blue). 
    \textbf{E}, Validation MSE during training. 
    \textbf{F}, Test direct relative $L^2$ error. 
    \textbf{G,H}, Pointwise error for a representative branch, showing reduced prediction error with warm-starting. 
    \textbf{I--L}, Gray--Scott evaluated on held-out parameters with untrained, pretrained, and warm-started initializations. 
    \textbf{I}, Mean number of distinct solutions per parameter (left axis) and reference-set coverage (right axis). 
    \textbf{J--L}, Solution-count maps over the $(\rho,\mu)$ plane for the three initializations.}
    \label{fig:ablation}
\end{figure}

Having established PatternFormer's ability to recover complex solution landscapes, we investigate the sources of this capability by comparing three deterministic training strategies under identical architectures, data and compute: (i) random initialization of the frozen LLM weights with LoRA adaptation~\cite{hu2022lora}, (ii) fine-tuning from the pretrained Qwen2.5-7B model, and (iii) warm-starting from an earlier stop-token PatternFormer checkpoint. This isolates the contributions of general LLM pretraining and of task-specific warm-starting from an already adapted scientific model.

We first evaluate the effect of LLM pretraining on Example~\ref{ex:1d1p} (Fig.~\ref{fig:ablation}A--D). Although this problem can be learned effectively with LoRA adaptation alone, the pretrained model converges faster and provides more accurate predictions at representative parameters. The benefit becomes more pronounced when transferring across equations. For Example~\ref{ex:2d}, warm-starting from the Example~\ref{ex:1d2p} checkpoint consistently reduces the validation loss compared with training a fresh adapter from the pretrained Qwen initialization (Fig.~\ref{fig:ablation}E). It also improves the test direct relative $L^2$ error, from a median of $1.1\times10^{-2}$ to $6\times10^{-3}$ (Fig.~\ref{fig:ablation}F), with substantially reduced pointwise error on representative solutions (Fig.~\ref{fig:ablation}G,H).

We further evaluate the three initialization strategies on the more challenging Gray--Scott system with $D_A{=}2.5\times10^{-4}$ and $D_S{=}5\times10^{-4}$. Under deterministic generation, the random-weight model (LoRA on randomly initialized backbone weights) recovers $1.44$ distinct steady states per parameter, the pretrained model recovers $3.23$, and the warm-started model reaches $5.18$ (Fig.~\ref{fig:ablation}I). Reference-set coverage increases correspondingly from $0.07$ to $0.22$ and $0.34$; coverage here is a relative check, since the Gray--Scott reference set is only a lower bound on an effectively unbounded family (Supplementary Section~4), so the distinct verified count is the primary measure. The solution-count maps over the $(\rho,\mu)$ plane show the same progressive improvement (Fig.~\ref{fig:ablation}J--L).

Together, these experiments reveal two complementary sources of performance. General LLM pretraining provides a useful initialization for learning scientific solution representations, while warm-starting from an already adapted PatternFormer model transfers problem-specific structure and further improves recovery of coexisting states. These findings suggest that scientific foundation models can accumulate transferable knowledge across related physical systems.
\section{Discussion}

PatternFormer introduces a new paradigm for scientific machine learning: learning the \emph{space of possible solutions} rather than predicting a single solution. Many physical systems, from nonlinear reaction--diffusion models in biology~\cite{kuang2015glioma} to pattern-forming reaction networks~\cite{kirshtein2020colon,rempala2017reduction}, are intrinsically multistable, yet existing neural operators are predominantly designed for single-valued mappings and therefore cannot directly represent the coexistence of multiple valid states. By formulating a parameter-to-solution-set map as an autoregressive generation problem, PF transforms multistable PDE solving from branch-by-branch discovery into direct solution-landscape generation. This shift enables a single model to recover complete families of coexisting steady states in one forward pass.

Across nonlinear elliptic boundary-value problems and the open-ended Gray--Scott Turing system, PF demonstrates this capability in two regimes: deterministic recovery of the complete branch set when the multiplicity is known, and open-ended generation of diverse physically consistent patterns---including states absent from the training data---when it is not. Machine learning is now applied across the sciences, from biomolecular property prediction~\cite{wang2020topnettree} to differential-equation solving, and these results suggest that foundation-model approaches can move beyond learning input--output functions toward learning structured scientific landscapes.

In this work, we realize this framework by adapting a pretrained general-purpose LLM (Qwen2.5-7B) with LoRA. Starting from this foundation model, PF is progressively adapted across nonlinear elliptic equations and the Gray--Scott system, spanning different dimensions, parameter spaces, and nonlinear structures. The same framework transfers across these problems through warm-starting rather than independent training, showing that PF is a reusable strategy rather than a task-specific network. Future extensions to broader PDE families may enable increasingly general solution-landscape models.  

Two directions offer natural opportunities for further improvement, each with a concrete path forward. First, purely feed-forward generation degrades far beyond the training parameter range; PF mitigates this by seeding homotopy continuation from its own predictions, which transports the complete solution set to distant parameters. Second, each model is trained on a single fixed spatial discretization; extending PF to heterogeneous meshes and data structures, toward discretization-agnostic generalization, is a natural next step.

\section{Methods}
\label{sec:methods}

\subsection{Set-valued operator formulation}
\label{subsec:formulation}

For a parametric steady reaction--diffusion problem with parameter
$p\in\mathcal{P}$, the governing equations may admit multiple coexisting steady solutions,
\[
\mathcal{S}(p)=\{u_1,\dots,u_{K(p)}\},\qquad u_k\in H^1(\Omega),
\]
where the cardinality $K(p)$ varies across parameter space. Conventional neural operators approximate single-valued mappings,
\[
p\mapsto u(p),
\]
and therefore cannot directly represent a parameter-to-solution-set relationship. We instead learn the set-valued operator
\[
\mathcal{G}: \mathcal{P}\rightarrow 2^{H^1(\Omega)},\qquad
p\mapsto\mathcal{S}(p),
\]
which returns the collection of coexisting steady states associated with a given parameter.

Because autoregressive models operate on ordered sequences rather than unordered sets, we convert each solution set into a canonical sequence. We introduce a monotone scalar functional $\Phi(u)$ to define a deterministic ordering: the signed integral
\[
\Phi(u)=\int_0^1u\,\mathrm{d}x
\]
for the one-dimensional problems and the spatial mean of the activator field,
\[
\Phi(A,S)=\langle A\rangle,
\]
for the Gray--Scott system. Rare ties are resolved using the field energy as a secondary criterion. The ordering permutation is therefore
\[
I(p)=\operatorname{argsort}
\big(\Phi(u_1),\dots,\Phi(u_{K(p)})\big),
\]
which converts the original unordered set into the canonical sequence
\[
\big(u_{I_1},u_{I_2},\dots,u_{I_{K(p)}}\big),
\]
satisfying
\[
\Phi(u_{I_1})\leq\cdots\leq\Phi(u_{I_{K(p)}}).
\]

This representation transforms set-valued operator learning into sequence generation. Similar canonical ordering strategies have been used in machine learning to convert unordered objects into ordered representations, including set-to-sequence learning~\cite{vinyals2016order}, set prediction~\cite{carion2020detr}, and object generation through language-model formulations~\cite{chen2022pix2seq}. Here we adapt this principle to the coexisting steady-state solution sets of multistable PDEs.

\subsection{Ground-truth data generation}
\label{subsec:datagen}

All reference solution sets are generated using classical multiple-solution solvers. Examples~\ref{ex:1d1p} and~\ref{ex:2d} are computed using the companion-based multilevel finite-element method~\cite{hao2024cbmfem}, Example~\ref{ex:1d2p} is generated using homotopy continuation with solution maps~\cite{hao2020spatial}, and Gray--Scott patterns are obtained using an efficient GPU-based quasi-Newton tensor-product solver~\cite{hao2025quasinewton}. 

For every problem, each stored solution is refined by Newton iteration until the residual satisfies
\[
\|F(u)\|<10^{-9}
\]
on the discretization used for subsequent evaluation. Trivial homogeneous states are excluded when counting coexisting nontrivial solutions.

For the square-domain problems (Example~\ref{ex:2d} and Gray--Scott), the domain possesses the dihedral symmetry group $D_4$. Solutions related by rotations or reflections correspond to the same physical pattern. We therefore reduce each solution set to a single representative from each $D_4$ symmetry orbit before training and evaluation. Details of the individual numerical pipelines are provided in the Supplementary Information.

\subsection{Model architecture}
\label{subsec:architecture}

PatternFormer is constructed by adapting a pretrained decoder-only large language model to operate on continuous scientific representations rather than text tokens. We use the open-source pretrained Qwen2.5-7B model~\cite{qwen2025}, with hidden dimension $3584$, and adapt it using low-rank adaptation (LoRA)~\cite{hu2022lora} with rank $r=16$ applied to the attention projections. All original LLM weights remain frozen.

The model receives parameters and solution representations as continuous vectors through six components (Fig.~\ref{fig:overview}):

\begin{enumerate}
\item A frozen convolutional autoencoder with encoder $\mathcal{E}$ and decoder $\mathcal{D}$, which maps a solution field to a latent representation
\[
z=\mathcal{E}(u)\in\mathbb{R}^{256}
\]
and reconstructs the field through
\[
u=\mathcal{D}(z).
\]

\item A learnable set of structural embeddings consisting of a block-start embedding $\mathbf{v}_{\mathrm{s}}$, a solution-separator embedding $\mathbf{v}$, and a stop embedding $\mathbf{v}_{\mathrm{p}}$. Each embedding has dimension $3584$, matching the input space of Qwen2.5-7B.

\item An input matching layer 
\[
g_{\mathrm{in}},
\]
implemented as a two-layer multilayer perceptron (MLP) with GELU activations, which maps the concatenated parameter--latent vector
\[
[z;p]
\]
into the LLM embedding space.

\item The pretrained Qwen2.5-7B transformer, which processes these continuous embeddings through its causal attention mechanism.

\item An output matching layer 
\[
g_{\mathrm{out}},
\]
which maps transformer hidden states back to the latent dimension.

\item A dual prediction head consisting of a latent regression head for generating solution representations and a marker classifier used for adaptive termination when the solution multiplicity is known.
\end{enumerate}

The text tokenizer and word embeddings are not used. The pretrained transformer therefore serves as a sequence-generation backbone for continuous scientific representations rather than as a language model operating on discrete tokens. In principle, the same framework can be transferred to other pretrained decoder architectures by retraining only the lightweight matching layers, structural embeddings, and adaptation modules.

\subsection{Sequence construction and training}
\label{subsec:sequence}

For a parameter $p$ with ordered solutions
$u_{(1)}\prec\cdots\prec u_{(K)}$, PF constructs a continuous sequence block
\[
B(p)=
\big(
\mathbf{v}_{\mathrm{s}},
g_{\mathrm{in}}(\mathbf{0},p),
\mathbf{v},
\tilde z_1,
\mathbf{v},
\tilde z_2,
\dots,
\mathbf{v},
\tilde z_K,
\mathbf{v}_{\mathrm{p}}
\big),
\]
where $\mathbf{v}_{\mathrm{s}}$ and $\mathbf{v}_{\mathrm{p}}$ denote the start and
termination markers, $\mathbf{v}$ denotes a solution marker, and
\[
\tilde z_k=g_{\mathrm{in}}([z_k,p]),\qquad
z_k=\mathcal{E}(u_{(k)}).
\]
The parameter is represented by the special case $z=\mathbf{0}$, allowing the
same projection mechanism to encode both parameters and solutions.

During training, the target parameter $p_\ast$ is preceded by a set of
in-context examples,
\begin{equation}
    \label{eq:input}
    B(p_1),\ldots,B(p_M),B(p_\ast),
\end{equation}
where each context block contains the complete ordered solution set associated
with its parameter. These examples provide the transformer with demonstrations
of the solution landscape. For the elliptic PDE examples, the context
parameters are sampled randomly from the training set, whereas for the more
complex Gray--Scott system they are selected locally around the target
parameter. During the final fine-tuning stage, the context is removed so that
the model learns to generate directly from the parameter alone.

PF is trained autoregressively with teacher forcing. Given previously generated
solution slots, the transformer predicts the latent representation of the next
solution,
\[
\hat z_k =
G_\theta(p,\hat z_1,\ldots,\hat z_{k-1}).
\]
All predictions are decoded into fields
$\hat u_k=\mathcal{D}(\hat z_k)$ and optimized using a combination of data
supervision and a physics residual.

For a predicted field $\hat u_k$, we define the residual loss
\[
R(\hat u_k;p)=
\frac{1}{N}
\sum_{i=1}^{N}
\left(
-\Delta_h\hat u_k(x_i)
+\mathcal{N}(\hat u_k(x_i);p)
-g(x_i)
\right)^2.
\]
Rather than minimizing this residual without constraint, which would favor
trivial low-residual states in some systems, we use a hinged residual penalty,
\[
\lambda_{\rm pde}
\operatorname{ReLU}(R(\hat u_k;p)-\epsilon),
\]
which encourages physically valid states while preventing the loss from
driving all outputs toward a trivial solution. The threshold $\epsilon$ is
chosen between the residual of valid steady states and the residual of
non-solution fields (See Supplementary Section~5 for more details).

\paragraph{Stop-token formulation.}

For problems where the number of coexisting solutions can be determined
theoretically, PF learns the solution count through a termination signal. After
the final ground-truth solution, a stop marker is appended and the model is
trained to identify this point.

The objective is
\begin{equation}
    \label{eq:loss-stop}
\mathcal{L}_{\rm stop}
=
\sum_{k=1}^{K(p)}
\left[
\lambda_{\rm mse}
\|\hat u_k-u_k\|^2
+
\lambda_{\rm pde}
\operatorname{ReLU}(R(\hat u_k;p)-\epsilon)
\right]
+
\lambda_{\rm ce}\mathcal{L}_{\rm marker},
\end{equation}
where $\mathcal{L}_{\rm marker}$ is the cross-entropy loss for predicting
solution and termination markers.

This formulation allows PF to generate exactly the required number of
solutions without specifying the output dimension in advance.

\paragraph{Fixed-budget design for Gray--Scott.} On Gray--Scott, whose coexisting set is effectively unbounded, this same head collapses to premature termination and emits only a few solutions, so there we fix the budget instead. The target block now holds a fixed number $K$ of solution slots ($K{=}24$), with no stop marker and the classification head disabled ($\lambda_{\text{ce}}{=}0$). Its first $K_t$ slots (the head), where $K_t=K(p)$ is the number of solutions provided for that parameter in the dataset, are teacher-forced on those reference solutions, while the remaining slots (the tail) free-run on the model's own predictions (Fig.~\ref{fig:overview}D):
\begin{equation}\label{eq:targetblock}
B(p_\star)=\mathbf{v}_{\mathrm{s}},\,p_\star,\ \underbrace{\mathbf{v}\,\tilde z_1,\,\dots,\,\mathbf{v}\,\tilde z_{K_t}}_{\substack{\text{teacher-forced head}\\ \text{(data loss)}}}\ \;\big|\;\ \underbrace{\mathbf{v}\,\hat z_{K_t+1},\,\dots,\,\mathbf{v}\,\hat z_{K}}_{\substack{\text{free-run tail}\\ \text{(physics residual)}}}.
\end{equation}
The loss accordingly applies field- and latent-space reconstruction losses on the head and the hinged physics residual throughout,
\begin{equation}\label{eq:loss}
\mathcal{L}=\sum_{k=1}^{K}\Big(\underbrace{\big(\lambda_{\text{mse}}\|\hat u_k-u_k\|^2+\lambda_{\text{lat}}\|\hat z_k-z_k\|^2\big)\,\mathbf{1}[k\le K_t]}_{\substack{\text{data loss:}\\ \text{field {+} latent MSE (head slots)}}}+\underbrace{\lambda_{\text{pde}}\, \operatorname{ReLU}\!\big(R(\hat u_k;p)-\epsilon\big)}_{\text{physics residual (all slots)}}\Big),
\end{equation}
so the hinged residual drives the tail to valid solutions beyond the data. Scheduled sampling counters exposure bias.

Because the number of coexisting solutions is heavily skewed across parameters (most carry few, the high-multiplicity ones are rare), in both cases we draw each parameter for an epoch with a probability that increases with its solution count $K_t$ (multiplicity-weighted sampling, the exact weighting given in Supplementary Section~5), so the rare high-multiplicity parameters are not neglected.

Each model is trained in three phases. We first pretrain the autoencoder on reconstruction alone and then freeze it (a UNet in one dimension, an MLP for the finite-element problem, and a convolutional network for Gray--Scott). We then train the full model with the in-context blocks present (the with-context phase). We finally fine-tune it with the context removed (the no-context phase), resumed from the with-context checkpoint, so that at test time the model generates from the target parameter alone.

\begin{algorithm}[t]
\caption{Training of the set-valued operator.}
\label{alg:train}
\begin{algorithmic}[1]
\Statex \textbf{Data.} A classical multi-solution solver returns the ordered coexisting set $\mathcal{S}(p){=}\{u_{(1)}\prec\cdots\prec u_{(K_t)}\}$ at each $p$; the frozen autoencoder gives latents $z_k=\mathcal{E}(u_{(k)})$.
\Statex \textbf{Training} (Qwen2.5-7B frozen; only LoRA, the projectors and the heads optimized).
\For{each epoch}
    \State Draw a target parameter $p_\star$ by multiplicity-weighted sampling (probability increasing with $K_t$).
    \State Form the input $B(p_1),\dots,B(p_M),\,B(p_\star)$ of Eq.~\ref{eq:input}, the context blocks $B(p_1),\dots,B(p_M)$ dropped in the no-context phase.
    \If{stop-token design ($K(p)$ known a priori)}
        \State Target block: $K_t$ slots then a stop marker; minimize $\mathcal{L}_{\text{stop}}$ (Eq.~\ref{eq:loss-stop}).
    \ElsIf{fixed-budget design (open-ended count, e.g.\ Gray--Scott)}
        \State Target block: $K$ slots with the head/tail split of Eq.~\ref{eq:targetblock}; minimize $\mathcal{L}$ (Eq.~\ref{eq:loss}).
    \EndIf
    \State Update LoRA, the projectors and the heads with AdamW.
\EndFor
\end{algorithmic}
\end{algorithm}

\subsection{Inference and post-processing}
\label{subsec:inference}

Where training teacher-forces the head on the ground truth, evaluation has none, so the whole block is filled by autoregressive generation. Conditioned on the query parameter $p_\star$, the model emits the solution latents in a single pass,
\begin{equation}\label{eq:autoreg}
\hat z_k=G_\theta\big(p_\star,\,\hat z_1,\dots,\hat z_{k-1}\big),\qquad k=1,\dots,K,
\end{equation}
each new solution conditioned on the parameter and on every solution already generated. This causal dependence is what a causal attention mask enforces: with $q_i,k_j$ the query and key at positions $i,j$, the masked self-attention logit is
\begin{equation}\label{eq:causal}
\ell_{ij}=\begin{cases}q_i^{\top}k_j/\sqrt{d},& j\le i,\\[2pt] -\infty,& j>i,\end{cases}
\end{equation}
so the slot that generates $\hat z_k$ attends only to the parameter and to $\hat z_1,\dots,\hat z_{k-1}$, never to later slots. The markers are fixed scaffolding, so the model never emits a discrete token, only continuous latents. Each predicted latent is decoded to a field $\hat u_k=\mathcal{D}(\hat z_k)$, re-encoded by $\mathcal{E}$ and fed back as the next input, which keeps the trajectory on the autoencoder's latent manifold.

A single such pass is deterministic and already returns a set of coexisting fields. Sharper accuracy is then optional and external. Every field $\hat u_k$ seeds the same classical solver used to generate the data, run to a residual $\lVert F\rVert<10^{-9}$, after which the converged fields are deduplicated up to the problem's discrete symmetry (relative $L^2$ below $0.15$) and the distinct ones counted. Relaxing the demanded tolerance removes this step entirely, since the raw outputs already clear the bar (Fig.~\ref{fig:tolerance}).

On the open-ended Gray--Scott set we can trade compute for additional steady states. We reuse the same autoregressive recursion (Eq.~\ref{eq:autoreg}) but add fresh Gaussian noise to each latent before it is fed back, so pass $t$ generates
\begin{equation}\label{eq:noise}
\hat z_k^{(t)}=G_\theta\big(p_\star,\,\hat z_1^{(t)},\dots,\hat z_{k-1}^{(t)}\big)+\sigma\,\xi_k^{(t)},\qquad \xi_k^{(t)}\sim\mathcal{N}(0,\mathbf{I}).
\end{equation}
Because each perturbed latent re-enters the context for the next step, the noise propagates along the autoregressive chain and steers the whole pass onto a neighbouring but self-consistent set of fields, which refinement then projects onto genuine steady states. Different noise draws explore different neighbours, so the deduplicated union $\bigcup_{t=1}^{T}\{\hat u_k^{(t)}\}_{k=1}^{K}$ over $T$ passes accumulates distinct steady states. Every added pass is one more forward evaluation (Fig.~\ref{fig:gs}D).

\subsection{Cross-PDE warm-starting}
\label{subsec:warmstart}

Rather than training each model independently, we investigate whether the
representation learned by PF transfers across different multistable systems.
Each new PDE is initialized from a previously adapted PF model and further
fine-tuned on the new equation.

The one-dimensional single-parameter problem
(Example~\ref{ex:1d1p}) is initialized from the general pretrained Qwen2.5-7B
model. The subsequent one-dimensional two-parameter problem
(Example~\ref{ex:1d2p}) is initialized from this adapted model, and the
two-dimensional elliptic problem (Example~\ref{ex:2d}) is initialized from the
Example~\ref{ex:1d2p} model. During transfer, the LoRA parameters, output
projection and prediction heads are reused, while the input projection is
reinitialized when the parameter dimension changes.

For Gray--Scott, we compare initialization from the general pretrained model
with initialization from an earlier Gray--Scott model using the stop-token
formulation. This isolates the benefit of transferring a PDE-adapted
representation when the generation strategy itself changes.

To distinguish transfer learning from simply training a larger model, we also
perform controlled comparisons in which the architecture, training data,
optimization procedure and computational budget are identical, but the
initialization is either random, pretrained-only, or warm-started from a
previously adapted PF model. These experiments quantify the contribution of
pretraining and PDE-specific adaptation to the recovered solution diversity.

\subsection{Evaluation metrics}
\label{subsec:metrics}

For Examples~\ref{ex:1d1p}--\ref{ex:2d} we report the relative $L^2$ error before and after Newton refinement, the number of Newton iterations and final residual, solution-count accuracy against the known multiplicity $K(p)$, and wall-clock cost. For Gray--Scott, which lacks a complete reference set, a pattern is accepted when $\lVert F\rVert<10^{-9}$, and we report the number of distinct converged states per parameter and the states generated beyond the reference data. All timing comparisons are made at equal accuracy ($\lVert F\rVert<10^{-9}$). Full metric definitions and the timing protocol are given in Supplementary Section~7, and the optimizer and hyperparameters in Supplementary Section~3.

\section*{Reporting summary}
Further information on research design is available in the Nature Portfolio Reporting Summary linked to this article.

\section*{Data availability}
The data analysed in this study consist of numerically generated PDE solution sets rather than empirical measurements. All solvers, configuration files and fixed random seeds needed to regenerate them are included in the public code repository (see Code availability), so every dataset can be reproduced in full. 

\section*{Code availability}

The complete source code for data generation, training, inference and evaluation---including all configuration files, fixed random seeds, and the scripts that regenerate every figure and reported statistic---is publicly available at \url{https://github.com/changzhipeng1-prog/PatternFormer} under an open-source (MIT) licence.

\bibliographystyle{unsrtnat}
\bibliography{references}

\section*{Acknowledgments}
Z.C. and W.H. were supported by National Institute of General Medical Sciences through grant 1R35GM146894. W.H. and W.Y. were supported by NSF DMS-2533995. W.H. was also supported by the Huck Chair in AI Mathematical Modeling from Penn State University's Huck Institutes of the Life Sciences.

\section*{Author contributions}
W.H. conceived the study. W.H. and Z.C. developed the methodology and designed the experiments. Z.C. implemented the code, performed the experiments, analysed the results and prepared the figures. Z.C. wrote the initial draft. W.H., W.Y. and Z.C. reviewed and edited the manuscript.

\section*{Competing interests}
The authors declare no competing interests.

\section*{Additional information}
Correspondence and requests for materials should be addressed to Wenrui Hao (wxh64@psu.edu).

\end{document}